\documentclass[letterpaper, 10 pt, conference]{ieeeconf}
\IEEEoverridecommandlockouts
\makeatletter
\let\NAT@parse\undefined
\makeatother

\usepackage{amsmath,amssymb}
\usepackage{mathtools}
\usepackage{dsfont}
\usepackage{xcolor}
\usepackage{booktabs,multirow}
\usepackage[flushleft]{threeparttable}

\usepackage{cite}

\usepackage{caption}
\usepackage{graphicx}
\graphicspath{{figures/}}

\usepackage[hidelinks]{hyperref}
\hypersetup{
	colorlinks=true,
	linkcolor=blue!60!black,
	citecolor=blue!60!black,
	urlcolor=blue!60!black,
}

\newtheorem{ass}{Assumption}
\newtheorem{thm}{Theorem}
\newtheorem{prop}{Proposition}
\newtheorem{lem}{Lemma}
\newtheorem{cor}{Corollary}
\newtheorem{rem}{Remark}

\title{\LARGE \bf
	Truncated backpropagation through time induces a two-sided gradient turnpike property
}

\author{Julian D. Schiller, Matthias A. Müller%
\thanks{This work was supported by the Deutsche Forschungsgemeinschaft (DFG, German Research Foundation) -- Project number: 535860958.} %
\thanks{Julian D. Schiller and Matthias A. Müller are with the Leibniz University Hannover, Institute of Automatic Control, 30167 Hannover, Germany.
	{\tt\small schiller@irt.uni-hannover.de},
	{\tt\small mueller@irt.uni-hannover.de}.}%
}

\begin{document}

\maketitle
\thispagestyle{empty}
\pagestyle{empty}

\begin{abstract}

Truncated backpropagation through time reduces the computational burden of training recurrent models by restricting gradient computations to shorter temporal contexts. This induces two distinct boundary effects: missing past information perturbs the forward state evolution, while truncating the backward recursion omits future information. This motivates decoupling the interval over which gradient contributions are averaged for the parameter update from the surrounding temporal context over which they are computed. Under exponential decay of the state-transition Jacobians and mild regularity conditions, this leads to a two-sided gradient turnpike property: the error between truncated- and full-context gradient contributions decays exponentially with the distance from both context boundaries. We derive bounds for the averaged update gradient, yielding guarantees on gradient alignment and approximate full-context stationarity; under unbiased interval sampling and a Polyak--Łojasiewicz condition, these further imply a bound on the resulting optimality gap. Numerical experiments on the Wiener--Hammerstein benchmark illustrate the role of past and future context and the trade-off between predictive performance and computational cost.

\end{abstract}

\section{Introduction}

Recurrent models are widely used across various domains, including dynamical systems, time-series forecasting, and sequence learning \cite{Pillonetto2025,Hewamalage2021,Lipton2015}. Their recurrent state enables them to retain information from the past and thereby model temporal dependencies over long horizons. Training such models, however, requires accounting for the dependence of the loss on the recurrent state trajectory. The standard approach is backpropagation through time (BPTT), which computes gradients by unrolling the recurrent dynamics over the entire training sequence \cite{Werbos1990}. For long sequences, full BPTT can become computationally and memory intensive, motivating truncated backpropagation through time (TBPTT), where gradients are computed using only shorter segments of the training sequence \cite{Williams1990,Sutskever2013,Beintema2023a}.

A central consequence of truncation is that gradient computations are affected by the restricted temporal window on which they are performed. On the one hand, the recurrent state at the beginning of such a window generally differs from the state obtained by propagating the model over the complete preceding sequence. This motivates the use of a burn-in phase, in which additional past samples are processed before the loss is evaluated \cite{Kapturowski2019,Bonassi2022,Schiller2025b}. Related approaches in nonlinear system identification reconstruct the initial state of a training subsequence from preceding input--output data \cite{Beintema2023a,Forgione2022}. On the other hand, the gradient contribution associated with a given time instant generally depends on losses occurring at later times through the backward recursion. Truncating this recursion at a fixed finite right boundary therefore also discards future loss information. Hence, truncated training naturally gives rise to two distinct boundary effects: one caused by missing past information in the forward state evolution and one caused by missing future information in the backward gradient computation.

Existing theoretical analyses have addressed these two effects largely separately. On the backward side, bounds on the bias introduced by truncating long-range gradient contributions in TBPTT are derived under a geometric decay condition and subsequently used to adapt the truncation horizon \cite{Aicher2020}. Related work obtains unbiased gradient estimators by randomizing the truncation length and appropriately reweighting the resulting gradient contributions \cite{Tallec2017}. On the forward side, contractive models imply that the influence of sufficiently remote past information decays with temporal distance \cite{Miller2019}, while our recent work shows how a finite burn-in phase can approximate full-sequence training under less restrictive finite-memory conditions \cite{Schiller2025b}. To the best of our knowledge, however, existing results do not \emph{jointly} quantify how missing past information in the forward path and missing future information in the backward path affect gradient computations for recurrent models. 
Importantly, these effects are coupled, as discrepancies in the forward state evolution also perturb the backward sensitivities.

The resulting two-sided structure is reminiscent of turnpike phenomena in finite-horizon optimal control and estimation, where optimal trajectories remain close to distinguished reference trajectories over most of the horizon, with deviations concentrated near the initial and terminal boundaries \cite{Faulwasser2022,Schiller2025,Trelat2015,Schiller2026a}. Such turnpike properties have been established for states, inputs, outputs, and adjoint variables, and corresponding exponential characterizations typically consist of terms that decay with the distance from the horizon boundaries.

In the same spirit, we show that an analogous two-sided phenomenon arises directly at the level of gradient contributions in the training of nonlinear recurrent models with truncated temporal context.
To capture the two boundary effects separately, we distinguish an update interval, over which gradient contributions are averaged to obtain the gradient used in the parameter update law, from a surrounding temporal context over which these contributions are computed.
Under exponential decay of the state-transition Jacobians and mild regularity conditions, we show that the error between truncated- and full-context gradient contributions decays exponentially with the distance from both truncation boundaries. 
Consequently, extending the context into the past or future independently tightens the corresponding contribution to the gradient-error bound.
This yields explicit bounds for the averaged gradient used in the parameter update law and, in turn, guarantees on gradient alignment and approximate full-context stationarity. 
Moreover, under unbiased interval sampling, these guarantees extend to the full training objective; combined with a Polyak--Łojasiewicz condition, this yields a bound on the optimality gap relative to the optimal value of the full-context loss.
The theoretical findings are illustrated on a standard nonlinear system identification benchmark, highlighting how past and future context affect the accuracy and computational efficiency of truncated learning.

\section{Problem setting}

This section introduces the learning problem and mathematical setup for the subsequent analysis.

\subsubsection*{Notation}
The set of non-negative integers is denoted by $\mathbb{N}$.
Intervals of time indices are understood as integer intervals.
The notation \(\|\cdot\|\) denotes the Euclidean norm for vectors and its induced matrix norm. The cardinality of a finite set $\mathcal A$ is denoted by $|\mathcal A|$.

\subsection{Full-context learning problem}
In this work, we consider the problem of learning the parameters $\theta\in\mathbb{R}^o$ of a general nonlinear discrete-time state-space model in the form of
\begin{align}
	x^+ &= f(x,u;\theta),\label{eq:sys_1}\\
	y &= h(x,u;\theta),\label{eq:sys_2}
\end{align}
where $x\in\mathbb{R}^n$ is the state of the system, $x^+\in\mathbb{R}^n$ is the successor state, $u\in\mathbb{R}^m$ is the input, and $y\in\mathbb{R}^p$ is the model output. The dynamics and output equation of the model are characterized by nonlinear functions $f$ and $h$, both of which are parameterized by $\theta$. This formulation encompasses standard recurrent neural network (RNN) models, including gated architectures such as LSTMs~\cite{Hochreiter1997} and GRUs~\cite{Cho2014}, as well as more modern sequence models such as Mamba~\cite{Gu2024}.

For training the model parameters $\theta$, we consider a given input sequence $(u_t)_{t\in\mathcal T}$ with time index set $\mathcal{T}:=[0,T-1]$, $T\in\mathbb{N}$, $T\geq 1$, and corresponding measured target outputs $(y_t^{\mathrm d})_{t\in\mathcal T}$.
For each $t\in\mathcal{T}$, let \(l_t(x;\theta)\) denote the per-time-step loss induced by the model output \(h(x,u_t;\theta)\) and the corresponding target \(y_t^d\).
We aim to minimize the full training loss
\begin{equation}\label{eq:loss}
	L(\theta) := \frac{1}{T}\sum_{t\in\mathcal{T}} l_t(x_t;\theta)
\end{equation}
evaluated along the full-context model simulation
\begin{equation}\label{eq:loss_states}
	x_0=0, \quad x_{t+1} = f(x_t,u_t;\theta), \quad  t\in\mathcal{T}.
\end{equation}
The initial state is fixed at zero for simplicity, as commonly done in the context of black-box system identification using RNNs; however, alternative initializations may also be considered.
The gradient of $L(\theta)$ can be decomposed as
\begin{equation}\label{eq:loss_grad}
	\nabla_\theta L(\theta) = \frac{1}{T}\sum_{t\in\mathcal{T}} g_t(\theta),
\end{equation}
where $g_t(\theta)$ is the contribution of the parameter dependence at time \(t\) to the full-context gradient, including its influence on subsequent losses through the dynamics.
These contributions can be expressed using the full-context backward sensitivity recursion (BPTT) introduced in Section~\ref{sec:analysis_preliminaries}.

Motivated by TBPTT, we introduce a \emph{gradient-update interval} $\mathcal{U}=[a,b]\subseteq\mathcal{T}$ and define the corresponding full-context \emph{update gradient} as
\begin{equation}\label{eq:gradient_full_context}
	{G}_{\mathcal{U}}(\theta) := \frac{1}{|\mathcal{U}|} \sum_{t\in\mathcal{U}}g_t(\theta);
\end{equation}
see Figure~\ref{fig:intervals} for an illustration.
The corresponding stochastic gradient descent (SGD) iterations take the form
\begin{equation}\label{eq:SGD}
	\theta_{i+1} = \theta_i -\eta_i G_{\mathcal{U}_i}(\theta_i), \quad i\in\mathbb{N},
\end{equation}
where $\mathcal{U}_i$ denotes a randomly sampled gradient-update interval and $\eta_i>0$ is the learning rate; compare, e.g., \cite{Bottou2018}.

\subsection{Truncating the context}
\begin{figure}
	\centering
	\vspace{2ex}\includegraphics[]{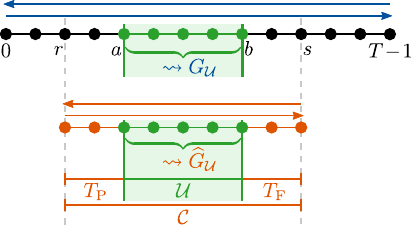}\vspace{1ex}
	\caption{
		Illustration of the gradient-update interval \(\mathcal U=[a,b]\) and the temporal context \(\mathcal C=[r,s]\). The full-context update gradient \(G_{\mathcal U}\) uses forward and backward recursions over the entire training sequence (blue), whereas the truncated-context update gradient \(\widehat G_{\mathcal U}\) uses recursions restricted to \(\mathcal C\), with past and future context extensions \(T_{\mathrm P}\) and \(T_{\mathrm F}\) (orange).
	}
	\label{fig:intervals}
\end{figure}

In the considered TBPTT scheme, both the forward model simulation and backward sensitivity recursion are restricted to a shorter temporal context. Consequently, the full-context update gradient used in~\eqref{eq:SGD} is replaced by a truncated-context approximation.
Hence, we associate with each interval $\mathcal U$ a truncated \emph{gradient-context interval} $\mathcal{C}=[r,s]\supseteq\mathcal{U}$ such that
\begin{equation}\label{eq:intervals}
	\emptyset \neq \mathcal{U}\subseteq\mathcal{C}\subseteq\mathcal{T}.
\end{equation}
In the following, we refer to $\mathcal{U}$ and $\mathcal{C}$ simply as the update and context intervals. Define
\begin{equation}\label{eq:TL_TR}
	T_\mathrm{P}:= a-r, \qquad   T_\mathrm{F} := s - b.
\end{equation}
Then, \(T_{\mathrm P}\) and \(T_{\mathrm F}\) denote the past- and future-context extensions, respectively; see Figure~\ref{fig:intervals}.
Now, let $\hat{g}_t(\theta)$ denote the local parameter-gradient contribution associated with time $t$, computed using an independently initialized forward simulation and a backward sensitivity recursion restricted to the context $\mathcal{C}$; compare Figure~\ref{fig:intervals}.
Averaging the contributions over $\mathcal{U}$ yields the truncated-context update gradient
\begin{equation}\label{eq:gradient_trunc_context}
	\widehat G_{\mathcal U}(\theta)
	:=\frac{1}{|\mathcal U|}\sum_{t\in\mathcal U}\hat g_t(\theta).
\end{equation}
At iteration \(i\in\mathbb{N}\), \(\widehat G_{\mathcal U_i}(\theta_i)\) replaces \(G_{\mathcal U_i}(\theta_i)\) in~\eqref{eq:SGD}.

\begin{rem}[Mini-batch SGD]
	Averaging truncated gradients over independently sampled update intervals in the SGD iterations~\eqref{eq:SGD} reduces sampling variance and permits parallel evaluation. The subsequent gradient-error bounds extend directly to mini-batch averages via the triangle inequality. Moreover, the truncated gradients can also be used with adaptive optimizers such as Adam~\cite{Kingma2015}; compare also \cite{Bottou2018}.
\end{rem}

The following analysis quantifies the discrepancies between the full- and truncated-context gradient contributions and their corresponding update gradients in~\eqref{eq:gradient_full_context} and~\eqref{eq:gradient_trunc_context}.

\section{Theoretical analysis}\label{sec:analysis}

Throughout this section, for brevity we consider a fixed input sequence \(u=(u_0,\ldots,u_{T-1})\) and a particular parameterization \(\theta\), and suppress algorithm iterations \(i\).
Consequently, the following assumptions and results are formulated for this fixed setting; they can be made uniform with respect to inputs and parameters if the required assumptions are imposed uniformly over all admissible input sequences and~parameters.

We first introduce the required machinery and assumptions and then analyze the gradient error and its implications for learning.
We indicate truncated-context quantities by a hat, while leaving their full-context counterparts undecorated.

\subsection{Preliminary definitions and main assumptions} \label{sec:analysis_preliminaries}

For the given input sequence $u$ and parameter $\theta$, let $(x_t)_{t\in\mathcal{T}}$ denote the full-context state sequence satisfying
\begin{equation}\label{eq:x_full}
	x_{t+1} = f(x_{t},u_t;\theta), \quad t\in\mathcal{T}, \quad x_0=0.
\end{equation}
For a context interval $\mathcal{C}=[r,s]\subseteq\mathcal{T}$, consider its truncated-context counterpart $(\hat{x}_t)_{t\in\mathcal{C}}$ such that
\begin{equation}\label{eq:x_trunc}
	\hat{x}_{t+1} = f(\hat{x}_{t},u_t;\theta), \quad t\in\mathcal{C}, \quad \hat{x}_{r} = 0.
\end{equation}

The zero initialization in~\eqref{eq:x_trunc} is adopted for simplicity.
This standard choice facilitates simple, numerically robust, and computationally efficient training and allows context intervals to be sampled and evaluated independently of their temporal order~\cite{Xu2023a,AllenZhu2019}.
Alternatively, zero initialization may be replaced by any \(\hat x_r\in\mathcal X\); the bounds derived below then depend on the initial-state mismatch \(\|\hat x_r-x_r\|\), with the past-context contribution vanishing if \(\hat x_r=x_r\).

Define the Jacobians evaluated along the two trajectories as
\begin{align*}
	&A_t := D_xf(x_t,u_t;\theta),  &&B_t := D_\theta f(x_t,u_t;\theta), &&t\in\mathcal{T},\\
	&\hat{A}_t := D_xf(\hat{x}_t,u_t;\theta),  &&\hat{B}_t := D_\theta f(\hat{x}_t,u_t;\theta), &&t\in\mathcal{C},
\end{align*}
with the corresponding transition Jacobians
\begin{align}
	&\Phi(t,k) := A_{t-1}A_{t-2}\cdots A_{k},
	\quad &&0\leq k<t\leq T, \ \label{eq:phi_t_full}\\
	&\hat{\Phi}(t,k) := \hat{A}_{t-1}\hat{A}_{t-2}\cdots \hat{A}_{k},
	\quad &&r\leq k<t\leq s, \label{eq:phi_t_trunc}
\end{align}
and $\Phi(k,k) = \hat{\Phi}(k,k) = I$.

To derive the gradient contributions $g_t$ and $\hat{g}_t$ in~\eqref{eq:gradient_full_context} and \eqref{eq:gradient_trunc_context}, respectively, consider the Lagrangian associated with the learning problem~\eqref{eq:loss}--\eqref{eq:loss_states}, omitting normalization:
\begin{equation}\label{eq:lagrangian}
	\mathcal{L}=\sum_{t\in\mathcal{T}}\left(l_t(x_t;\theta)+p_{t+1}^\top(f(x_t,u_t;\theta)-x_{t+1})\right).
\end{equation}
The adjoint variables (or \emph{backward sensitivities}) $p_t$ are determined by imposing $\nabla_{x_t}\mathcal{L}=0$ for all $t\in[1,T]$, yielding the full-context backward recursion
\begin{equation}
	p_T = 0,
	\quad 
	p_t = \nabla_xl_t(x_t;\theta) + A_t^\top p_{t+1},
	\quad
	t\in \mathcal{T}. \label{eq:pt_full}
\end{equation}
Since the system dynamics~\eqref{eq:x_full} and adjoint recursion~\eqref{eq:pt_full} are satisfied, we have $T\nabla_\theta L(\theta)=\nabla_\theta\mathcal L$, yielding
\begin{equation*}
	T\nabla_\theta L(\theta) = \sum_{t\in\mathcal{T}}\left(\nabla_\theta l_t(x_t;\theta) + B^\top_tp_{t+1}\right).
\end{equation*}
Thus, the local gradient contributions $g_t$ in~\eqref{eq:gradient_full_context} are given by
\begin{equation}\label{eq:gt_full}
	g_t := \nabla_\theta l_t(x_t;\theta)+B_t^\top p_{t+1}, \quad t\in\mathcal{T}.
\end{equation}
The corresponding truncated-context expressions follow by constructing the analogous Lagrangian on \(\mathcal C=[r,s]\), yielding
\begin{align}
	\hat{p}_{s+1} = 0, \quad
	\hat{p}_t = \nabla_xl_t(\hat{x}_t;\theta) + \hat{A}_t^\top \hat{p}_{t+1}, \quad
	t\in \mathcal{C} \label{eq:pt_trunc}
\end{align}
and
\begin{equation}
	\hat{g}_t := \nabla_\theta l_t(\hat{x}_t;\theta)+\hat{B}_t^\top \hat{p}_{t+1}, \quad t\in\mathcal{U}. \label{eq:gt_trunc}
\end{equation}
Note that the backward recursion~\eqref{eq:pt_trunc} incorporates losses throughout the context $\mathcal{C}$, including the $T_\mathrm{F}$ future losses $l_{b+1},...,l_s$, whereas only the contributions $\hat{g}_t$ for $t\in\mathcal{U}$ are averaged to form the gradient $\widehat{G}_\mathcal{U}$ in~\eqref{eq:gradient_trunc_context}; compare Figure~\ref{fig:intervals}.

We now impose the main assumptions on the setup.

\begin{ass}[Compact state domain] \label{ass:X_set}
	There exists $\mathcal{X}\subset\mathbb{R}^n$, convex and compact, such that $x_t\in\mathcal{X}$ for all $t\in\mathcal{T}$ and $\hat{x}_t\in\mathcal{X}$ for all $t\in\mathcal{C}$.
\end{ass}

\begin{ass}[Decaying state-transition Jacobians]
	\label{ass:exp_decay_jac}
	There exist $C\geq 1$ and $\rho\in(0,1)$ such that
	\begin{equation}\label{eq:ass:exp_decay_jac}
		\bigl\|D_x\phi_{t,k}(x)\bigr\|
		\leq C\rho^{t-k}
	\end{equation}
	for all $0\leq k\leq t\leq T$ and all $x\in\mathcal{X}$, where $\phi_{t,k}(x)$ denotes the $(t-k)$-step flow of system~\eqref{eq:sys_1} initialized at $x$ and evolving under the parameter $\theta$ given the corresponding input sequence $(u_k,\ldots,u_{t-1})$.
\end{ass}

Assumption~\ref{ass:exp_decay_jac} requires exponential decay of the state-transition Jacobians evaluated along a model trajectory. Note that this property allows individual one-step Jacobians to be expansive. Specifically, it implies exponential forgetting of initial-state differences under a common input sequence, which is relevant for recurrent models whose dependence on remote past information decays over time. Related properties are standard in the literature; compare, e.g., \cite{Beintema2023a,Schiller2025b,Pillonetto2025}.

\begin{ass}[Bounded{\,}and{\,}Lipschitz{\,}derivatives]
	\label{ass:derivatives}
	\ There exist $\bar{l},\bar{B}>0$ such that $\|\nabla_xl_t(x;\theta)\|\leq \bar{l}$ for all $t\in\mathcal{T}$ and $\|D_\theta f(x,u_t;\theta)\|\leq \bar{B}$ for all $t\in\mathcal{U}$, for all $x\in\mathcal{X}$.
	Moreover, there exist $L_A,L_B,L_x,L_\theta>0$ such that
	\begin{align*}
		\|D_xf(x,u_t;\theta) - D_xf(\tilde{x},u_t;\theta)\| &\leq L_A\|x-\tilde{x}\|,
		&& t\in\mathcal{C}\\
		\|D_\theta f(x,u_t;\theta) - D_\theta f(\tilde{x},u_t;\theta)\| &\leq L_B\|x-\tilde{x}\|,
		&& t\in\mathcal{U},\\
		\|\nabla_x l_{t}(x;\theta)-\nabla_x l_{t}(\tilde{x};\theta)\| &\leq L_x\|x-\tilde{x}\|,
		&& t\in\mathcal{C},\\
		\|\nabla_\theta l_{t}(x;\theta)-\nabla_\theta l_{t}(\tilde{x};\theta)\| &\leq L_\theta\|x-\tilde{x}\|,
		&& t\in\mathcal{U}
	\end{align*}
	for all $x,\tilde{x}\in\mathcal{X}$.
\end{ass}

Assumption~\ref{ass:derivatives} involves standard regularity requirements. They follow, e.g., when the model dynamics and loss functions are sufficiently smooth on $\mathcal{X}$, with bounded input and target sequences.

\subsection{Gradient turnpike}

We first show that our setup introduced in the previous section implies uniform boundedness of the backward sensitivities introduced in~\eqref{eq:pt_full} and \eqref{eq:pt_trunc}. 

\begin{lem}\label{lem:pt_bound}
	Consider an arbitrary context interval $\mathcal{C}\subseteq\mathcal{T}$.
	Let Assumptions \ref{ass:exp_decay_jac} and \ref{ass:derivatives} hold. Then, the backward sensitivities $p_t$ and $\hat{p}_t$ are uniformly bounded on their corresponding time intervals. Specifically, there exists $\bar{p}>0$ such that
	\begin{equation}\label{eq:lem_pt_bound}
		\|p_t\|\leq \bar{p},\ t\in \mathcal{T}, \qquad
		\|\hat{p}_t\|\leq \bar{p},\ t\in\mathcal{C}.
	\end{equation}
	Moreover, these bounds extend to \(t=T\) and \(t=s+1\), respectively.
\end{lem}

\begin{proof}
	We start by unrolling the recursion in~\eqref{eq:pt_full} over $[t,T-1]$ for some $t\in\mathcal{T}$. Since $p_T=0$, this yields
	\begin{equation*}
		p_t = \sum_{k=t}^{T-1}\Phi(k,t)^\top \nabla_x l_k(x_k;\theta), \quad t\in\mathcal{T},
	\end{equation*}
	where $\Phi$ is from \eqref{eq:phi_t_full}. Note that $\Phi(k,t) = D_x\phi_{k,t}(x_t)$ for all $t,k\in\mathcal{T}$ with $k\geq t$.
	Thus, from the triangle inequality, Assumptions~\ref{ass:exp_decay_jac} and \ref{ass:derivatives}, and the geometric series, we obtain that
	\begin{equation}
		\|p_t\|
		\leq \sum_{k=t}^{T-1}C\bar{l}\rho^{k-t}
		\leq \frac{C\bar{l}}{1-\rho}, \quad t\in\mathcal{T}. \label{eq:proof_lem}
	\end{equation} 
	Thus, letting $\bar{p}:={C\bar{l}}/({1-\rho})$ establishes the first property in~\eqref{eq:lem_pt_bound}; the second one directly follows from similar arguments applied to~\eqref{eq:pt_trunc}.
	
	Finally, the terminal conditions \(p_T=0\) and \(\hat p_{s+1}=0\) in~\eqref{eq:pt_full} and~\eqref{eq:pt_trunc} imply that the bounds in~\eqref{eq:lem_pt_bound} also hold for \(t=T\) and \(t=s+1\), respectively, completing the proof.
\end{proof}

The following result establishes exponential decay of the gradient discrepancy away from both context boundaries.

\begin{thm}[Pointwise gradient turnpike]\label{thm:gradient_turnpike}
	\ Suppose that Assumptions~\ref{ass:X_set}--\ref{ass:derivatives} hold for all admissible pairs of update and context intervals $(\mathcal U,\mathcal C)$, with constants independent of the interval choices.
	Then, there exist constants \(K_{\mathrm P},K_{\mathrm F}>0\), independent of \(\mathcal U\) and \(\mathcal C\), such that
	\begin{equation}\label{eq:gradient_turnpike}
		\|\hat{g}_t(\theta) - g_t(\theta)\| \leq K_\mathrm{P}\rho^{t-r} +  K_\mathrm{F}\rho^{s-t},\quad t\in\mathcal{U},
	\end{equation}
	with $\rho\in(0,1)$ from Assumption~\ref{ass:exp_decay_jac}.
\end{thm}

\begin{proof}
	Consider arbitrary context and update intervals $\mathcal{C}$ and $\mathcal{U}$ satisfying~\eqref{eq:intervals}.
	Define the forward state error $e^x_t:= \hat{x}_t - x_t$, $t\in\mathcal{C}$, where we recall that ${x}_t$ and $\hat{x}_t$ satisfy~\eqref{eq:x_full} and \eqref{eq:x_trunc}, respectively.
	Define $x(\tau) := \hat{x}_r +\tau(x_r-\hat{x}_r)$, $\tau\in[0,1]$.
	Since $x_r,\hat{x}_r\in\mathcal{X}$ by Assumption~\ref{ass:X_set} with $\mathcal{X}$ being convex, it follows that $x(\tau)\in\mathcal{X}$ for all $\tau\in[0,1]$.
	Using the Fundamental Theorem of Calculus, we obtain
	\begin{align*}
		e^x_t:= \hat{x}_t - x_t
		&= \phi_{t,r}(x(0))-\phi_{t,r}(x(1))\\
		&= -\int_{0}^{1}D_x\phi_{t,r}(x(\tau))x_rd\tau.
	\end{align*}
	Taking norms, using the triangle inequality, and applying~\eqref{eq:ass:exp_decay_jac} then leads to
	\begin{equation}
		\|e^x_{t}\| \leq \int_{0}^{1}\|D_x\phi_{t,r}(x(\tau))\|\|\hat{x}_r-x_r\|d\tau\leq C\bar{x}\rho^{t-r} \label{eq:proof_forward_stability}
	\end{equation}
	for all $t\in\mathcal{C}$, using~\eqref{eq:x_trunc} and $\bar{x}:=\max_{x\in\mathcal{X}}\|x\|<\infty$.
	
	Define the backward sensitivity error $e^p_t = \hat{p}_t - p_t$, $t\in[r,s+1]$.
	By using~\eqref{eq:pt_full} and \eqref{eq:pt_trunc}, we obtain, for each $t\in\mathcal{C}$,
	\begin{align}
		e^p_t
		&= \nabla_xl_t(\hat{x}_t;\theta) + \hat{A}_t^\top \hat{p}_{t+1} - \nabla_xl_t(x_t;\theta) - A_t^\top p_{t+1}\nonumber \\
		&= A_t^\top e^p_{t+1} + \xi_t, \label{eq:proof_backward_step}
	\end{align}
	with
	\begin{equation}
		\xi_t :=  \nabla_xl_t(\hat{x}_t;\theta) - \nabla_xl_t(x_t;\theta)
		+ (\hat{A}_t - A_t)^\top \hat{p}_{t+1}.
	\end{equation}
	Taking norms, using the triangle inequality, Lipschitz-continuity of $\nabla_xl_t$ and $D_xf$ from Assumption~\ref{ass:derivatives}, uniform boundedness of $\|\hat{p}_{t}\|\leq \bar{p}$ from Lemma~\ref{lem:pt_bound}, and the definition of the errors $e^x_t$ and $e^p_t$ yields, for each $t\in\mathcal{C}$,
	\begin{align*}
		\|\xi_t\|
		&\leq  \|\nabla_xl_t(\hat{x}_t;\theta) - \nabla_xl_t(x_t;\theta)\|
		+ \|\hat{A}_t - A_t\|\| \hat{p}_{t+1}\|\\
		&\leq L_x\|e^x_t\| + L_A\bar{p}\|e^x_t\|.
	\end{align*}
	Hence, using~\eqref{eq:proof_forward_stability} and defining $C_\xi := C\bar{x}(L_x + L_A\bar{p})$ leads to
	\begin{equation} \label{eq:proof_bound_xi}
		\|\xi_t\| \leq C_\xi\rho^{t-r}, \quad t\in\mathcal{C}.
	\end{equation}
	
	Unrolling~\eqref{eq:proof_backward_step} to the context boundary $s$ and using~\eqref{eq:phi_t_full} and $\hat{p}_{s+1}=0$ from~\eqref{eq:pt_trunc} yields
	\begin{align}\label{eq:proof_ep_t}
		e^p_t = -\Phi(s+1,t)^\top p_{s+1} + \sum_{k=t}^s\Phi(k,t)^\top\xi_k, \quad t\in\mathcal{C}.
	\end{align}
	Here, by using submultiplicativity of the norm, Assumption~\ref{ass:exp_decay_jac}, and Lemma~\ref{lem:pt_bound}, the first term on the right-hand side satisfies
	\begin{equation}\label{eq:proof_ep_t_term1}
		\|\Phi(s+1,t)^\top p_{s+1}\| \leq C\bar{p}\rho^{s+1-t}.
	\end{equation}
	The second term in~\eqref{eq:proof_ep_t} satisfies, by the same arguments together with application of \eqref{eq:proof_bound_xi} and the geometric series,
	\begin{align}
		\left\|\sum_{k=t}^s\Phi(k,t)^\top\xi_k\right\|
		&\leq \sum_{k=t}^s\|\Phi(k,t)\|\|\xi_k\|\nonumber\\
		&\leq CC_\xi\rho^{t-r}\sum_{k=t}^s\rho^{2(k-t)} \leq \frac{CC_\xi}{1-\rho^2}\rho^{t-r}.\label{eq:proof_ep_t_term2}
	\end{align}
	Then, from~\eqref{eq:proof_ep_t} with \eqref{eq:proof_ep_t_term1} and \eqref{eq:proof_ep_t_term2}, we can infer that
	\begin{equation}\label{eq:proof_res_p_1}
		\|e^p_t\| \leq \frac{CC_\xi}{1-\rho^2}\rho^{t-r} + C\bar{p}\rho^{s+1-t}, \quad t\in\mathcal{C}.
	\end{equation}
	We claim that the bound in~\eqref{eq:proof_res_p_1} also holds for $t=s+1\notin\mathcal{C}$ as well. 
	This is required below to cover the special case where $b=s$, that is, when the right boundaries of $\mathcal{U}$ and $\mathcal{C}$ coincide. Indeed, since $\hat{p}_{s+1}=0$ by~\eqref{eq:pt_trunc} and $\|p_{s+1}\|\leq \bar{p}$ by Lemma~\ref{lem:pt_bound}, it follows that $\|e^p_{s+1}\| = \|\hat{p}_{s+1} - p_{s+1}\| \leq \bar{p}$.
	Hence, the right-hand side of \eqref{eq:proof_res_p_1} evaluated at $t=s+1$ provides a trivial upper bound for $\|e^p_{s+1}\|$, since $C\geq 1$.
	Let $K_\mathrm{P}^p := CC_\xi/(1-\rho^2)$ and $K_\mathrm{F}^p := C\bar{p}$.
	Hence, we obtain
	\begin{equation}\label{eq:proof_res_p}
		\|e^p_t\| \leq K_\mathrm{P}^p\rho^{t-r} + K_\mathrm{F}^p\rho^{s+1-t}, \quad t\in[r,s+1].
	\end{equation}
	
	Now, consider the difference of the gradient contributions in \eqref{eq:gt_full} and \eqref{eq:gt_trunc} for $t\in\mathcal{U}$:
	\begin{equation*}
		\hat{g}_t-g_t = \nabla_\theta l_t(\hat{x}_t;\theta)+\hat{B}_t^\top \hat{p}_{t+1} -  \nabla_\theta l_t(x_t;\theta)-B_t^\top p_{t+1}.
	\end{equation*}
	Adding $0=B_t^\top \hat{p}_{t+1}-B_t^\top \hat{p}_{t+1}$ to the right-hand side, rearranging, and substituting the backward error $e_t^p$ leads to
	\begin{align*}
		\hat{g}_t-g_t =&\ \nabla_\theta l_t(\hat{x}_t;\theta)-  \nabla_\theta l_t({x}_t;\theta)\\
		&\ + B_t^\top e^p_{t+1}
		+(\hat{B}_t-B_t)^\top \hat{p}_{t+1}.
	\end{align*}	
	Since \(x_t,\hat x_t\in\mathcal X\) by Assumption~\ref{ass:X_set}, after taking norms and leveraging the triangle inequality, we can apply Assumption~\ref{ass:derivatives} and Lemma~\ref{lem:pt_bound} to further obtain that
	\begin{align}
		\|\hat{g}_t-g_t\|
		&\leq \|\nabla_\theta l_t(\hat{x}_t;\theta)-  \nabla_\theta l_t(x_t;\theta)\|\nonumber\\
		&\quad\, + \|B_t\|\| e^p_{t+1}\| +\|\hat{B}_t-B_t\|\|\hat{p}_{t+1}\| \nonumber\\
		&\leq (L_\theta+L_B\bar{p})\|e^x_t\| + \bar{B}\|e^p_{t+1}\| \label{eq:proof_gt}
	\end{align}
	for all $t\in\mathcal{U}$. Applying \eqref{eq:proof_forward_stability} and \eqref{eq:proof_res_p} to~\eqref{eq:proof_gt} then leads to
	\begin{equation*}
		\|\hat{g}_t-g_t\|
		\leq ((L_\theta+L_B\bar{p})C\bar{x} + \bar{B}K_\mathrm{P}^p\rho)\rho^{t-r} + \bar{B}K_\mathrm{F}^p\rho^{s-t}.
	\end{equation*}
	Thus, letting $K_\mathrm{P}:=(L_\theta+L_B\bar{p})C\bar{x} + \bar{B}K_\mathrm{P}^p\rho$ and $K_\mathrm{F}=\bar{B}K_\mathrm{F}^p$ establishes~\eqref{eq:gradient_turnpike}. Noting that these constants are uniform over all admissible $\mathcal{C}$ and $\mathcal{U}$ completes the proof.
\end{proof}

Theorem~\ref{thm:gradient_turnpike} establishes a two-sided gradient turnpike: the error between the truncated and full-context gradient contributions decays exponentially with the distance from either boundary of \(\mathcal C\). The past-boundary term results from the initialization of the truncated forward dynamics, whereas the future-boundary term reflects the truncation of the backward adjoint recursion. Consequently, the truncated-context gradient contributions closely approximate their full-context counterparts away from both boundaries.

The following two corollaries translate the gradient turnpike property established in Theorem~\ref{thm:gradient_turnpike} into direct design implications.

\begin{cor}[Gradient-contribution error]\label{cor:local_gradient} 
	Suppose that the conditions of Theorem~\ref{thm:gradient_turnpike} hold. Then, the gradient contributions in~\eqref{eq:gradient_turnpike} satisfy
	\begin{align*}
		\max_{t\in\mathcal{U}}\|\hat{g}_t(\theta) - g_t(\theta)\| \leq K_\mathrm{P}\rho^{T_\mathrm{P}} + K_\mathrm{F} \rho^{T_\mathrm{F}}.\\[-2ex]
	\end{align*}
\end{cor}

\begin{proof}
	Recall~\eqref{eq:TL_TR}; $t\in\mathcal{U}$ implies that $t-r\geq a-r = T_\mathrm{P}$ and $s-t\geq s-b = T_\mathrm{F}$.
	Thus,~\eqref{eq:gradient_turnpike} directly yields
	\begin{align*}
		\|\hat{g}_t-g_t\| \leq K_\mathrm{P}\rho^{T_\mathrm{P}} + K_\mathrm{F}\rho^{T_\mathrm{F}}, \quad t\in\mathcal{U}.
	\end{align*}
	Taking the maximum over $t \in\mathcal{U}$ completes the proof.
\end{proof}

\begin{cor}[Update gradient error]\label{cor:update_gradient} 
	Suppose the conditions of Theorem~\ref{thm:gradient_turnpike} hold.
	Then, the full- and truncated-context update gradients in~\eqref{eq:gradient_full_context} and \eqref{eq:gradient_trunc_context} satisfy
	\begin{equation}\label{eq:cor_gradient_update}
		\|\widehat{G}_\mathcal{U}(\theta) - G_\mathcal{U}(\theta)\| \leq \epsilon_G(T_\mathrm{P},T_\mathrm{F},|\mathcal{U}|),
	\end{equation}
	where
	\begin{equation}\label{eq:epsilon_G}
		\epsilon_G(T_\mathrm{P},T_\mathrm{F},|\mathcal{U}|) := \frac{1}{1-\rho}\cdot\frac{K_\mathrm{P}\rho^{T_\mathrm{P}}+K_\mathrm{F}\rho^{T_\mathrm{F}}}{|\mathcal{U}|}.
	\end{equation}
\end{cor}

\begin{proof}
	Under the conditions of Theorem~\ref{thm:gradient_turnpike}, the gradient turnpike property in~\eqref{eq:gradient_turnpike} applies.
	Thus, the update gradients in~\eqref{eq:gradient_full_context} and \eqref{eq:gradient_trunc_context} satisfy
	\begin{align*}
		\|\widehat{G}_\mathcal{U}(\theta) - G_\mathcal{U}(\theta)\|
		&\leq \frac{1}{|\mathcal{U}|}\sum_{t\in\mathcal{U}}\|\hat{g}_t(\theta) - g_t(\theta)\|\\
		&\leq \frac{1}{|\mathcal{U}|}\sum_{t\in\mathcal{U}}(K_\mathrm{P}\rho^{t-r} +  K_\mathrm{F}\rho^{s-t}).
	\end{align*}
	The application of~\eqref{eq:TL_TR} yields
	\begin{equation*}
		\|\widehat{G}_\mathcal{U}(\theta) - G_\mathcal{U}(\theta)\|
		\leq \frac{K_\mathrm{P}}{|\mathcal{U}|}\rho^{T_\mathrm{P}}\sum_{t\in\mathcal{U}}\rho^{t-a} +  \frac{K_\mathrm{F}}{|\mathcal{U}|}\rho^{T_\mathrm{F}}\sum_{t\in\mathcal{U}}\rho^{b-t}.
	\end{equation*}
	Recall that $\mathcal{U}=[a,b]$. Thus, using the geometric series, we obtain that
	\begin{equation*}
		\|\widehat{G}_\mathcal{U}(\theta) - G_\mathcal{U}(\theta)\|
		\leq \frac{K_\mathrm{P}}{|\mathcal{U}|}\rho^{T_\mathrm{P}}\frac{1}{1-\rho} +  \frac{K_\mathrm{F}}{|\mathcal{U}|}\rho^{T_\mathrm{F}}\frac{1}{1-\rho},
	\end{equation*}
	which establishes~\eqref{eq:cor_gradient_update} under \eqref{eq:epsilon_G}, completing the proof.
\end{proof}

Corollaries~\ref{cor:local_gradient} and~\ref{cor:update_gradient} directly characterize the dependence of the gradient-error bounds on the past- and future-context extensions \(T_\mathrm{P}\) and \(T_\mathrm{F}\). Increasing either context length exponentially reduces the corresponding error contribution. For fixed context extensions, increasing the update-interval length further reduces the error bound for the averaged update gradient: the boundary-layer contributions are diluted through averaging, yielding a decay of order \(\mathcal O(1/|\mathcal U|)\). Thus, \(T_\mathrm{P}\), \(T_\mathrm{F}\), and \(|\mathcal U|\) provide complementary design parameters for balancing gradient accuracy against computational effort.

\subsection{Implications for learning}

In this section, we investigate the implications of gradient-turnpike behavior for learning. Based on the magnitude of the truncated-context gradient \(\widehat G_{\mathcal U}\), we distinguish between regimes away from and close to stationarity. The following proposition relates these two regimes to the corresponding full-context gradient. Under an additional assumption on the sampling of the update intervals in~\eqref{eq:SGD}, the subsequent corollary relates the expected truncated-context update gradient to the gradient of the full loss and, under a Polyak--Łojasiewicz condition, provides a bound on the optimality gap.

\begin{prop}[Gradient alignment and stationarity]\label{prop:gradient_alignment}
	For an arbitrary update interval $\mathcal U\subseteq\mathcal T$, suppose that
	\begin{equation}\label{eq:corr_cond}
		\|\widehat G_{\mathcal U}(\theta)-G_{\mathcal U}(\theta)\|
		\leq \epsilon_G
	\end{equation}
	for some $\epsilon_G>0$. Then, the following statements hold:
		
	\begin{enumerate}
		\item If $\|\widehat G_{\mathcal U}(\theta)\|>\epsilon_G$, then
			\begin{equation*}
				0\leq \angle\bigl(\widehat{G}_{\mathcal U}(\theta),G_{\mathcal U}(\theta)\bigr) 
				\leq \arcsin\left( \frac{\epsilon_G}{\|\widehat G_{\mathcal U}(\theta)\|} \right) <\frac{\pi}{2}.
			\end{equation*}
		In particular, $\widehat G_{\mathcal U}(\theta)$ and $G_{\mathcal U}(\theta)$ are positively aligned.
		
		\item If $\|\widehat G_{\mathcal U}(\theta)\|\leq\delta$ for some $\delta\geq0$, then
		\begin{equation*}
			\|G_{\mathcal U}(\theta)\| \leq \delta+\epsilon_G.
		\end{equation*}
		In particular, truncated-context stationarity ($\delta=0$) implies full-context $\epsilon_G$-stationarity.
	\end{enumerate}
\end{prop}

\begin{proof}
	In the following, we omit the arguments $\theta$ and $\mathcal U$ of $G_{\mathcal U}(\theta)$ and $\widehat G_{\mathcal U}(\theta)$ for brevity.
	
	First, suppose that $\|\widehat{G}\|>\epsilon_G$, where $\epsilon_G>0$ corresponds to the bound in condition~\eqref{eq:corr_cond}. Then,
	\begin{align}
		\widehat{G}^\top {G}
		&= \|\widehat{G}\|^2 +\widehat{G}^\top({G}-\widehat{G}) \geq \|\widehat{G}\|^2 - \|\widehat{G}\|\|G-\widehat{G}\|\nonumber\\
		&\geq \|\widehat{G}\| \bigl(\|\widehat{G}\|-\epsilon_G\bigr)>0. \label{eq:proof_GTG}
	\end{align}
	Since $\widehat{G}^\top {G} = \|\widehat{G}\|\|G\|\cos\angle\bigl(\widehat{G},G\bigr)$, the strict inequality in \eqref{eq:proof_GTG} yields $\cos\angle\bigl(\widehat{G},G\bigr) > 0$, which further implies that
	\begin{equation}\label{eq:proof_GTG_phi}
		0\leq\angle\bigl(\widehat{G},G\bigr)<\pi/2,
	\end{equation}
	i.e., $\widehat{G}$ and $G$ are positively aligned.
	
	To tighten the angle bound in~\eqref{eq:proof_GTG_phi}, consider the unit vector $v:={G}/\|{G}\|$, noting that $\|{G}\|\neq0$ due to \eqref{eq:proof_GTG}. Let $P_\perp :=I-vv^\top$ denote the orthogonal projector onto the subspace perpendicular to $v$.
	Since $P_\perp G=0$, $P_\perp \widehat{G} = -P_\perp(G-\widehat{G})$.
	Taking norms, using that an orthogonal projection does not increase the norm, and applying~\eqref{eq:corr_cond} yields
	\begin{equation}\label{eq:proof_GTG_2}
		\|P_\perp \widehat{G}\| = \|P_\perp({G}- \widehat{G})\|
		\leq \|G- \widehat{G}\|
		\leq \epsilon_G.
	\end{equation}
	Moreover, $\sin\angle\bigl(\widehat{G},G\bigr) = \sin\angle\bigl(\widehat{G},v\bigr) = {\|P_\perp \widehat{G}\|}/{\|\widehat{G}\|}$, so that \eqref{eq:proof_GTG_2} leads to
	\begin{equation}\label{eq:proof_GTG_3}
		\sin\angle\bigl(\widehat{G},G\bigr)\leq {\epsilon_G}/{\|\widehat{G}\|}.
	\end{equation}
	Due to $\|\widehat{G}\|>\epsilon_G$, we have $0<{\epsilon_G}/{\|\widehat{G}\|}<1$.
	Together with~\eqref{eq:proof_GTG_phi} and the strict monotonicity of $\sin$ on $[0,\pi/2]$, this implies $\angle\bigl(\widehat{G},G\bigr) \leq \arcsin({\epsilon_G}/{\|\widehat{G}\|})$, which establishes the first statement.
	
	For the second statement, suppose that $\|\widehat G\|\leq\delta$ for some $\delta\geq0$. The triangle inequality and~\eqref{eq:corr_cond} directly give
	\begin{equation*}
		\|G\| \leq \|\widehat G\|+\|G-\widehat G\| \leq \delta+\epsilon_G,
	\end{equation*}
	which completes the proof.
\end{proof}

The proposition distinguishes two regimes relative to the truncation-error bound $\epsilon_G$. If $\|\widehat G_{\mathcal U}(\theta)\|>\epsilon_G$, the truncated-context gradient is positively aligned with the corresponding full-context gradient, with an explicit bound on their angular deviation. If, on the other hand, $\|\widehat G_{\mathcal U}(\theta)\|\leq\delta$, then the full-context gradient satisfies $\|G_{\mathcal U}(\theta)\|\leq\delta+\epsilon_G$, implying approximate full-context stationarity for sufficiently small $\delta$.

Whenever a certified truncation-error bound \(\epsilon_G\) is available, for example from Corollary~\ref{cor:update_gradient}, the applicable regime can be determined during optimization from \(\|\widehat G_{\mathcal U}\|\), which is already computed as part of the gradient update.

The preceding results hold for arbitrary update intervals, without assumptions on their sampling in the SGD iterations~\eqref{eq:SGD}. To translate the guarantees of Proposition~\ref{prop:gradient_alignment} to the full training loss, we now additionally assume that the sampling scheme is unbiased, i.e., such that
\begin{equation}\label{eq:unbiased_full_gradient}
	\mathbb{E}_{\mathcal U}\bigl[G_{\mathcal U}(\theta)\bigr] = \nabla_\theta L(\theta).
\end{equation}
Note that this is a common requirement in stochastic-gradient methods; see, e.g.,~\cite[Sec.~4]{Bottou2018}.
Condition~\eqref{eq:unbiased_full_gradient} is satisfied whenever each time index receives the same expected normalized sampling weight, i.e.,
\begin{equation*}
	\mathbb E_{\mathcal U}\left[
	\frac{\mathds{1}_{\{t\in\mathcal U\}}}{|\mathcal U|}
	\right]
	=\frac{1}{T},
	\qquad t\in\mathcal T,
\end{equation*}
where $\mathds{1}_{\{t\in\mathcal U\}}$ denotes the indicator function of the event $t\in\mathcal U$.
This holds, for example, when equal-length blocks forming a partition of \(\mathcal T\) are sampled uniformly.

\begin{cor}[Expected gradients]
	Let the sampling condition~\eqref{eq:unbiased_full_gradient} hold, with each sampled update interval satisfying condition~\eqref{eq:corr_cond} for a uniform $\epsilon_G>0$.
	Then,
	\begin{equation}\label{eq:expected_gradient_bound}
		\| \mathbb{E}_{\mathcal U}[\widehat G_{\mathcal U}(\theta)] -\nabla_\theta L(\theta) \| \leq \epsilon_G.
	\end{equation}
	Consequently, the statements of Proposition~\ref{prop:gradient_alignment} hold with \(\widehat G_{\mathcal U}(\theta)\) and \(G_{\mathcal U}(\theta)\) replaced by \(\mathbb E_{\mathcal U}[\widehat G_{\mathcal U}(\theta)]\) and \(\nabla_\theta L(\theta)\), respectively.
	
	If, additionally, $V^*:=\inf_\theta L(\theta)>-\infty$ and $L$ satisfies the Polyak--{\L}ojasiewicz condition
	\begin{equation}\label{eq:PL:cond}
		\frac{1}{2}\|\nabla_\theta L(\theta)\|^2 \geq \mu(L(\theta)-V^*)
	\end{equation}
	for some $\mu>0$, then
	\begin{equation}\label{eq:optimality_gap}
		L(\theta)-V^* \leq \frac{(\delta+\epsilon_G)^2}{2\mu}
	\end{equation}
	for any $\delta\geq0$ such that $\|\mathbb E_{\mathcal U}[\widehat G_{\mathcal U}(\theta)]\|\leq \delta$.
\end{cor}

\begin{proof}
	Leveraging condition~\eqref{eq:unbiased_full_gradient}, linearity of expectation, Jensen's inequality, and~\eqref{eq:corr_cond}, we obtain
	\begin{align*}
		\|\mathbb{E}_{\mathcal U}[\widehat G_{\mathcal U}(\theta)]-\nabla_\theta L(\theta) \|
		&= \|\mathbb{E}_{\mathcal U}[\widehat G_{\mathcal U}(\theta)-G_{\mathcal U}(\theta)]\|\\
		&\leq \mathbb{E}_{\mathcal U}[\|\widehat G_{\mathcal U}(\theta)-G_{\mathcal U}(\theta)\|]
		\leq\epsilon_G,
	\end{align*}
	which establishes~\eqref{eq:expected_gradient_bound}.
	Thus, the arguments used in the proof of Proposition~\ref{prop:gradient_alignment} directly apply with $\widehat G_{\mathcal U}$ and $G_{\mathcal U}$ replaced by $\mathbb{E}_{\mathcal U}[\widehat G_{\mathcal U}]$ and $\nabla_\theta L$, respectively.
	
	Now let $\|\mathbb E_{\mathcal U}[\widehat G_{\mathcal U}(\theta)]\|\leq \delta$ for some $\delta\geq0$, $V^*:=\inf_\theta L(\theta)>-\infty$ and suppose the PL condition in~\eqref{eq:PL:cond} holds. Using \eqref{eq:expected_gradient_bound} and the triangle inequality yields
	\begin{align*}
		L(\theta)-V^*
		&\leq \frac{1}{2\mu}\|\nabla_\theta L(\theta)\|^2\\
		&\leq \frac{1}{2\mu}(\|\mathbb{E}_{\mathcal U}[\widehat G_{\mathcal U}]\| + \|\mathbb{E}_{\mathcal U}[\widehat G_{\mathcal U}] - \nabla_\theta L(\theta)\|)^2\\
		&\leq \frac{(\delta+\epsilon_G)^2}{2\mu},
	\end{align*}
	which establishes~\eqref{eq:optimality_gap} and thus completes the proof.
\end{proof}

Corollary~3 yields a similar two-regime interpretation for the full training loss. If \(\|\mathbb{E}_{\mathcal U}[\widehat G_{\mathcal U}(\theta)]\|>\epsilon_G\), then \(-\mathbb{E}_{\mathcal U}[\widehat G_{\mathcal U}(\theta)]\) is a strict descent direction for \(L\) at \(\theta\). If instead \(\|\mathbb{E}_{\mathcal U}[\widehat G_{\mathcal U}(\theta)]\|\leq\delta\), then \(\|\nabla_\theta L(\theta)\|\leq\delta+\epsilon_G\); in particular, stationarity of the expected truncated-context gradient implies \(\epsilon_G\)-stationarity of the full training loss \(L\).

The Polyak--Łojasiewicz condition in~\eqref{eq:PL:cond} can hold for certain nonconvex objectives and is standard in the analysis of gradient-based optimization methods~\cite{Karimi2016}; however, it is not generally guaranteed for nonlinear training problems.
We emphasize that the preceding gradient-error, alignment, and stationarity guarantees are independent of this condition.
In the present setting, it allows translating approximate full-context stationarity into a bound on the global optimality gap $L(\theta)-V^*$. Specifically, for \(\delta=0\),
\begin{equation*}
	L(\theta)-V^* \leq \frac{\epsilon_G^2}{2\mu}.
\end{equation*}
According to Corollary~\ref{cor:update_gradient}, increasing the past- or future-context length, or increasing the update-interval length while keeping \(T_{\mathrm P}\) and \(T_{\mathrm F}\) fixed, reduces \(\epsilon_G\) and thereby tightens the preceding guarantees.

\section{Numerical example}
\label{sec:numerical}

We illustrate the theoretical results on the Wiener--Hammerstein benchmark \cite{Schoukens2026}. The dataset comprises a single input--output trajectory with a predefined split of 80,000 training and 20,000 validation samples. We identify a vanilla RNN model with 16 states described by
\begin{equation*}
	x^+=\tanh(Ax+Bu+b),\quad y=Cx+Du+d,
\end{equation*}
where the weighting matrices $A, B, C, D$ and bias vectors $b,d$ are learned by minimizing the squared output simulation error using $l_t(x_t;\theta)=\|Cx_t+Du_t+d-y_t^d\|^2$. Implementation details and code for reproducing the subsequent experiments are provided in the accompanying public repository\footnote{Code is available at \url{https://github.com/alesco-research/tbptt-gradient-turnpikes}}.

\begin{figure}
	\includegraphics[width=\columnwidth]{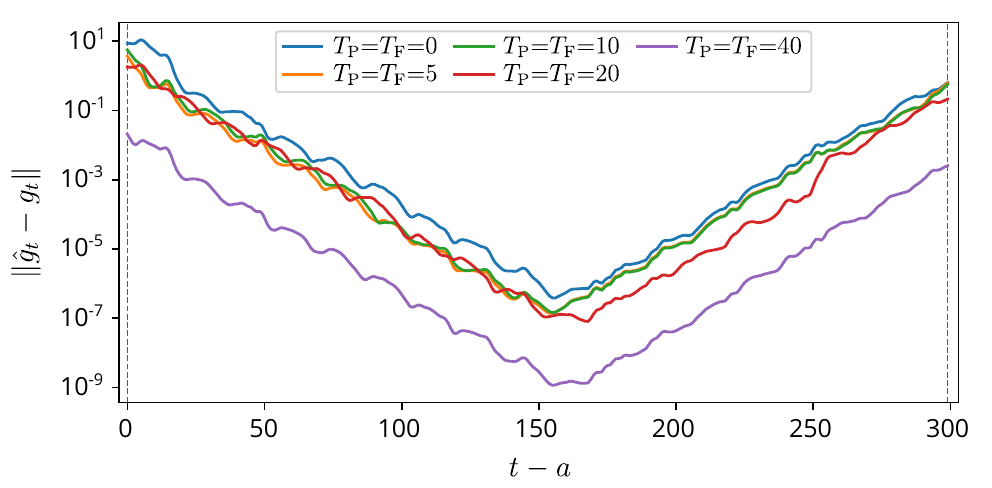}
	\caption{
		Pointwise discrepancy between the full and truncated-context gradient contributions evaluated over a fixed interval $\mathcal{U}=[a,b]$ for symmetric context lengths. Increasing the context length exponentially attenuates the boundary effects and thereby extends the interior region in which the truncated-context gradient contributions closely approximate their full-context counterparts.
		}
	\label{fig:gt}
\end{figure}

We first illustrate the truncation effects characterized by Theorem~\ref{thm:gradient_turnpike} and Corollary~\ref{cor:update_gradient}.
Since the results hold pointwise for arbitrary parameters \(\theta\), we consider five fixed reference parameterizations obtained using BPTT for different random seeds.
We then evaluate truncated-context gradients for \(T_{\mathrm P},T_{\mathrm F}\in\{0,5,10,20,40\}\).
Figure~\ref{fig:gt} illustrates the pointwise discrepancy $\|\hat g_t(\theta)-g_t(\theta)\|$ for different symmetric context lengths $T_{\mathrm P}=T_{\mathrm F}$, using one of the five reference parameterizations and a fixed evaluation interval $\mathcal U=[a,b]$ of length 300.
For short contexts, pronounced errors occur close to both boundaries, whereas increasing the context yields an extended low-error interior region. This is in line with Theorem~\ref{thm:gradient_turnpike} and directly reflects the two terms on the right-hand side of \eqref{eq:gradient_turnpike}: the errors induced at the left and right context boundaries decay exponentially into the interior.

\begin{figure}
	\includegraphics[width=\columnwidth]{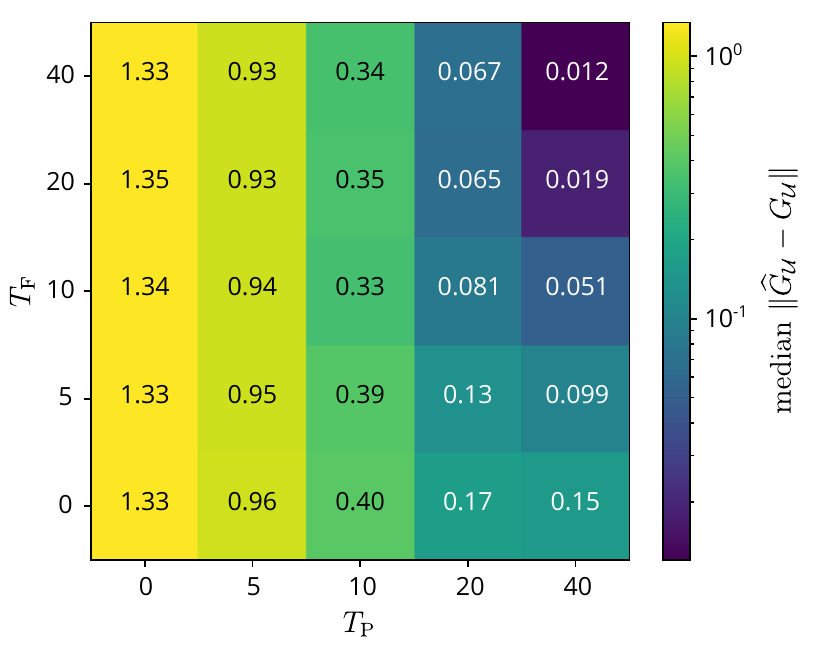}
	\caption{
		Median absolute discrepancy between the full and truncated-context update gradients as a function of the past and future context lengths. Values are aggregated over 20 common update intervals for each of the five reference parameterizations. Larger two-sided context substantially improves the full-context-gradient approximation.
		}
	\label{fig:GT}
\end{figure}

Figure~\ref{fig:GT} shows the median update-gradient discrepancy $\|\widehat G_{\mathcal U}(\theta)-G_{\mathcal U}(\theta)\|$ over the full $(T_{\mathrm P},T_{\mathrm F})$ grid. For each pair, the five reference parameterizations are evaluated on the same 20 update intervals, and the median is taken over the resulting 100 gradient discrepancies.
Increasing the past context consistently reduces the gradient discrepancy, while the benefit of extending the future context becomes particularly pronounced once the forward truncation error has been sufficiently reduced. This can be attributed to the fact that the forward and backward recursions are coupled: errors in the forward trajectory also perturb the backward sensitivities.
Overall, the median error decreases from 1.33 for \((T_{\mathrm{P}},T_{\mathrm{F}})=(0,0)\) to 0.012 for \((40,40)\), i.e., by a factor of approximately 110, consistent with Corollary~\ref{cor:update_gradient}.

To complement the preceding results, we numerically assess the contraction mechanism underlying Assumption~\ref{ass:exp_decay_jac} by estimating $\max_{k\in\mathcal K,x\in[-1,1]^{16}} \|D_x\phi_{k+20,k}(x)\|_2$.
To this end, we evaluate this norm for all $k\in\mathcal{K}:=[0,10^5-20]$ and sampled initial states, followed by multistart local optimization over the initial states for the largest candidates. Across the five reference parameterizations, the largest value found is 0.9708, providing numerical support for Assumption~\ref{ass:exp_decay_jac}.

\begin{table}[t]
	\centering
	\begin{threeparttable}
		\caption{
			Training and test performance of full-context BPTT and TBPTT with symmetric context $T_\mathrm{e}\!:=\! T_{\mathrm P}\!=\! T_{\mathrm F}$. Reported RMSEs are medians over five seeds; values in parentheses are median seed-paired RMSE ratios relative to BPTT; speedups are median seed-paired training-time~ratios.
		}
		\label{tab:training_comparison}
		\begin{tabular*}{\columnwidth}{@{\extracolsep{\fill}}lcccc@{}}
			\toprule
			Method
			& $T_\mathrm{e}$
			& Train RMSE
			& Test RMSE
			& Speedup \\
			\midrule
			BPTT
			& full
			& 0.02364\;(1.00$\times$)
			& 0.02284\;(1.00$\times$)
			& 1.00$\times$ \\
			TBPTT
			& 0
			& 0.03950\;(1.55$\times$)
			& 0.03850\;(1.58$\times$)
			& 12.99$\times$ \\
			TBPTT
			& 10
			& 0.03307\;(1.50$\times$)
			& 0.03248\;(1.51$\times$)
			& 5.71$\times$ \\
			TBPTT
			& 20
			& 0.04591\;(2.25$\times$)
			& 0.04527\;(2.25$\times$)
			& 5.64$\times$ \\
			TBPTT
			& 40
			& 0.02691\;(1.11$\times$)
			& 0.02622\;(1.12$\times$)
			& 5.68$\times$ \\
			\bottomrule
		\end{tabular*}
		\begin{tablenotes}[flushleft]
			\footnotesize
			\item BPTT uses $|\mathcal U|=1,000$; TBPTT uses $|\mathcal U|=50$ with mini-batch size 4. All models use 5,000 Adam updates and validation-based model selection.
			The separate benchmark test data comprise 78,800 samples, with the test RMSE computed after an initialization phase of 50 samples according to the benchmark convention \cite{Schoukens2026}.
		\end{tablenotes}
	\end{threeparttable}
\end{table}

Finally, in a separate training experiment, we investigate the practical trade-off between truncation and training performance for a fixed training budget of 5,000 optimizer updates and compare it to a challenging full-context BPTT benchmark.
The results are reported in Table~\ref{tab:training_comparison}, using the model checkpoint achieving the lowest validation loss among those periodically evaluated during training. Test performance is assessed on a separate test sequence used only for final evaluation.
For $T_{\mathrm P}{\,=\,}T_{\mathrm F}{\,=\,}40$, TBPTT approaches the BPTT reference closely, with median training and test RMSEs only 1.11 and 1.12 times those of the seed-paired BPTT reference, respectively, while reducing the training time by a factor of approximately 5.7. Hence, a sufficiently large two-sided context can retain much of the predictive performance of full BPTT while substantially reducing the computational effort.
Here, we observe that the training performance is not monotone in the context length: in particular, $T_{\mathrm P}{\,=\,}T_{\mathrm F}{\,=\,}20$ performs worse than $T_{\mathrm P}{\,=\,}T_{\mathrm F}{\,=\,}10$. Note that this does not contradict the preceding analysis, which characterizes the approximation of the full-context gradient at a fixed parameterization, but rather reflects the finite-budget, stochastic, and nonconvex nature of the resulting optimization problem.

\section{Conclusion}

In this paper, we showed that training nonlinear recurrent models with truncated temporal context, as in TBPTT, naturally gives rise to a two-sided gradient turnpike property.
Specifically, under exponential decay of state-transition Jacobians and mild regularity conditions, the discrepancy between truncated- and full-context gradient contributions decays exponentially away from both truncation boundaries, reflecting missing past information in the forward path and missing future information in the backward path. The resulting bounds quantify how omitted past and future context, as well as the update-interval length, affect the averaged gradient used for parameter updates, yielding guarantees on gradient alignment, approximate full-context stationarity, and, under an additional Polyak--Łojasiewicz condition, a bound on the corresponding optimality gap. Numerical results on the Wiener--Hammerstein benchmark illustrate the two-sided gradient turnpike property and show that TBPTT with a suitably chosen temporal context can substantially reduce computational effort while retaining performance close to full-context BPTT training.

\section*{Acknowledgment}
OpenAI ChatGPT was used during the preparation of this manuscript to assist with language refinement throughout the paper, literature discovery, the discussion and verification of mathematical derivations, and implementation and verification of numerical experiments. All AI-assisted content was critically reviewed, verified, and revised.


\end{document}